\documentclass{elsart}
\pdfoutput=1
\usepackage{amsmath,color}
\usepackage{mathrsfs}
\usepackage{graphicx}
\usepackage{natbib}
\usepackage{amssymb} 
\begin{document}
\begin{frontmatter}
\title{Bifurcation Phenomena in Two-Dimensional Piecewise Smooth
Discontinuous Maps} \author[cts]{Biswambhar Rakshit},
\author[rut]{Manjul Apratim}, \author[stan]{Parag Jain},
\author[eeiit]{Soumitro Banerjee\corauthref{cor}}
\corauth[cor]{Corresponding author.}
\ead{soumitro@ee.iitkgp.ernet.in} \address[cts]{Department of
Mathematics and Centre for Theoretical Studies, Indian Institute of
Technology, Kharagpur-721302, India} \address[rut]{Department of
Physics and Astronomy, Rutgers, The State University of New Jersey, NJ
08854-8019 USA} \address[stan]{ Department of Mechanical Engineering,
Stanford University, CA 94305, USA} \address[eeiit]{Department of
Electrical Engineering, Indian Institute of Technology,
Kharagpur-721302, India}

\begin{abstract}
In recent years the theory of border collision bifurcations has been
developed for piecewise smooth maps that are continuous across the
border, and has been successfully applied to explain nonsmooth
bifurcation phenomena in physical systems. However, many switching
dynamical systems have been found to yield two-dimensional piecewise
smooth maps that are discontinuous across the border. The theory for
understanding the bifurcation phenomena in such systems is not
available yet. In this paper we present the first approach to the
problem of analysing and classifying the bifurcation phenomena in
two-dimensional discontinuous maps, based on a piecewise linear
approximation in the neighborhood of the border. We explain the
bifurcations occurring in a physical system -- the static VAR
compensator used in electrical power systems, using the theory
developed in this paper. This theory may be applied similarly to other
systems that yield 2-D discontinuous maps.
\end{abstract}
\end{frontmatter}

\section{Introduction}
In recent years the bifurcations occurring in switching dynamical
systems have been the subject of great interest, as it is known that
many physical, engineering and biological systems actually embody
continuous-time evolution punctuated by discrete switching
events. Under stroboscopic sampling these switching dynamical systems
in general yield maps that have a discontinuity in the derivative
along subspaces or ``borderlines'' that divide the phase space into
two or more compartments. In these piecewise smooth (PWS) maps a new
type of bifurcation, called border collision bifurcation
\cite{nusse92,nusse95,mariobook,zhusubaliyev-book} occurs when a fixed
point collides with a borderline, resulting in a sudden change in the
Jacobian matrix. Most recent studies on border collision bifurcations
have been done on piecewise smooth maps that are continuous across the
borderline \cite{nusse95,feigin-mario,pre2d,cas2d}.

It has been reported that many switching dynamical systems yield maps
that not only have a discontinuity in the derivative, but also a
discontinuity in the function itself.  Discontinuity in the map arises
if, in the $n$ dimensional Poincar\'e section there exists an $(n-1)$
dimensional manifold such that infinitesimally close points at the two
sides of the manifold map to points that are far apart.  A few
electronic circuits exhibiting one-dimensional discontinuous maps was
given in \cite{discontcircuit}. The bifurcation theory for such
one-dimensional discontinuous maps have been developed
\cite{parag03,higham}, and have been successfully applied in analyzing
bifurcations in physical systems \cite{missedswitch}. However, in
recent years there has been evidence that many physical systems yield
two-dimensional (2D) discontinuous maps. For example, the
classical impact oscillator yields a 2D discontinuous map
if the impacting surface moves following a nonsmooth function
\cite{Budd-phyD01}. The static VAR controller used in electrical power
systems \cite{dobson1,dobson2} also has its dynamics given by a 2D
discontinuous map. 
 It has also been found that the dynamics of spiking
bursting activities of real biological neurons
\cite{Shil-phyA01,Rulf-pre01} as well as some business cycle models
\cite{Shushko_csf01} can be represented by 2D piecewise
smooth discontinuous maps. Therefore, to explain the bifurcation
phenomena in such systems, it is necessary to have a bifurcation
theory for 2D piecewise smooth discontinuous maps.

The bifurcation theory for 1D and 2D continuous piecewise smooth maps
is well developed \cite{1d98,cas2d}. The results related to the
existence of period-1 and period-2 orbits in general $n$-dimensional
PWS maps \cite{feigin-mario} have also been applied in practical
systems. Bifurcation theory for 1D discontinuous maps has been
reported \cite{parag03}. In the context of general $n$-dimensional
discontinuous maps, some important results related to the existence of
period-1 orbits \cite{higham} and period-2 orbits
\cite{partha-nld2007} have been published recently. For the special
case of $n=2$, these theories can be applied to obtain the conditions
of existence of period-1 and period-2 fixed points. However in the
context of a specific dynamical system, one is often interested in the
asymptotically stable behavior, which cannot be inferred from the
available theory. It has to be obtained through the analysis of the
existence and stability of periodic orbits and their stable and
unstable manifolds. Our investigations along this line showed that the
two-dimensional discontinuous map is a source of incredibly rich
dynamical behavior, created by complicated interactions between the
stable and unstable manifolds of fixed points and the line of
discontinuity. In this paper we report some results of this
investigation, and we hope that these will pave way for further
investigation on the dynamics of two-dimensional discontinuous maps.

The paper is organized as follows. In Section~2 we present an example
of a physical system that yields a discontinuous map, and illustrate
the peculiar dynamics exhibited by this system. As in the study of
continuous piecewise smooth systems, we use the piecewise linear
normal form representing the behavior of the system in the
neighborhood of the border-crossing fixed point, which is introduced
in Section~3. Then in Section~4 we develop a classification of the
bifurcation scenarios observed in this map, depending on the type of
the fixed points at the two sides of the border.

\section{A Physical Example: The Reactive Power Compensator \label{var}}

\begin{figure}[htb]
\begin{centering}
\includegraphics[width=.75\columnwidth]{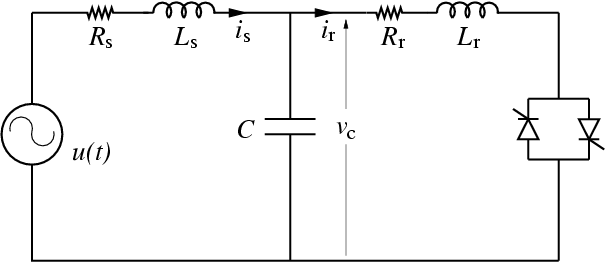}
\par\end{centering}
\caption{The Static Var Compensator Circuit. The parameters are 
$\omega=2\pi60$~rad/s, $L_{S}=0.195$~mH, $R_{s}=0.9$~m$\Omega$,
$L_{r}=1.66$~mH, $R_{r}=31.3$~m$\Omega$ and $C=1.5$~mF.
\label{fig:svar}}
\end{figure}

 The system that we shall consider is the static VAR
compensator used in electrical power systems, shown in
Fig~\ref{fig:svar}. It is an inductor-capacitor combination ($L_r$ and
$C$) in which
the reactive power consumed by the inductor is controlled by
back-to-back connected thyristor switches.  The system is connected to
a sinusoidal source $u(t)$ at the input (representing the rest of the
power system) through a transmission line of inductance $L_s$ and
resistance $R_s$. The two switches operate in the alternate
half-cycles, and are turned on by applying pulses at a phase angle
$\alpha$ with respect to the sinusoidal input $u(t)=\sin(\omega
t)$. They turn off when the current through the respective switches
reach a zero value.
\begin{figure}[htb]
\centering \includegraphics[width=0.76\columnwidth]{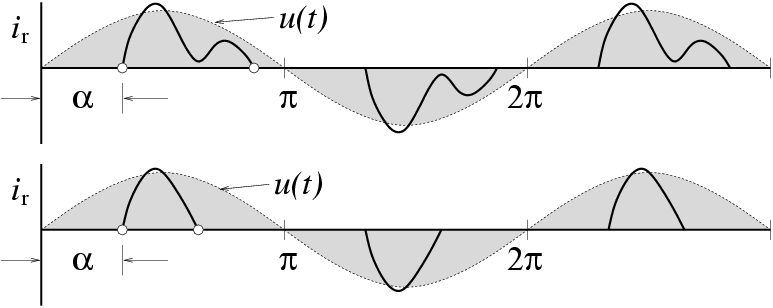}
\caption{The $i_r$ waveforms for two different initial conditions in
  $v_c$ and $i_s$.}
\label{svar-wave}
\end{figure}
The system is described by the state vector $\mathbf{x}(t)$ of the current 
$i_{r}(t)$, capacitor voltage $v_{c}(t)$ and the source current
$i_{s}(t)$. The dynamical equation during the {\sc on} state of either
switch is
\begin{equation}
\mathbf{\dot{x}}=\mathbf{A}_{\rm ON}\mathbf{x}+\mathbf{B}_{\rm ON}u
\label{eq:on}
\end{equation}
where:
\begin{equation}
\mathbf{A}_{\rm ON}\!=\!\left(\!\begin{array}{ccc}
-R_{r}/L_{r} & 1/L_{r} & 0\\
-1/C & 0 & 1/C\\
0 & -1/L_{s} & -R_{s}/L_{s}\end{array}\!\right) \!,\:\:\:
\mathbf{B}_{\rm ON}\!=\!\left(\!\begin{array}{c}
0\\
0\\
1/L_{s}\end{array}\!\right)
\label{eq:on-mat}
\end{equation}
During the {\sc off} phase, $i_{r}(t)$ is zero, and so the state vector is
two dimensional, given by ${\bf y}=[v_c \;\; i_s]^T$. The
corresponding dynamical equations are
\begin{equation}
\mathbf{\dot{y}}=\mathbf{A}_{\rm OFF}\mathbf{y}+\mathbf{B}_{\rm OFF}u
\label{eq:off}
\end{equation}
where
\begin{equation}
\mathbf{A}_{\rm OFF}=\left(\!\begin{array}{cc}
0 & 1/C\\
-1/L_{s} & -R_{s}/L_{s}\end{array}\!\right),\:\:\:\:\: \mathbf{B}_{\rm
  OFF}=
\left(\!\begin{array}{c}
0\\
1/L_{s}\end{array}\!\right).
\label{eq:off-mat}
\end{equation}

We obtain the discrete map by observing the state vector
stroboscopically, at each positive zero crossing of the input
sinusoid. Since the stroboscopic observations are made when the
switches are off, the discrete model must be two-dimensional. In this
system, the switching logic imposes a discontinuity in the Poincar\'e
map. To illustrate, consider the waveforms of the inductor current
$i_r$ in Fig.~\ref{svar-wave}. Suppose the switch is turned on at a
phase angle $\alpha$ and the waveform of the current $i_r$ is as
shown. At a slightly different initial condition, the point of dip may
reach a zero value, and so the instant of switch-off discontinuously
changes as shown in the figure. If for a specific initial condition of
$v_c$ and $i_s$ the waveform just grazes the $x$-axis, then initial
conditions at the two sides of this critical value map to widely
separated points in the phase space. This implies that the resulting
map is discontinuous. For a detailed derivation of the map, refer to
\cite{dobson1,dobson2}.

\begin{figure}[htb]
\includegraphics[width=0.45\columnwidth]{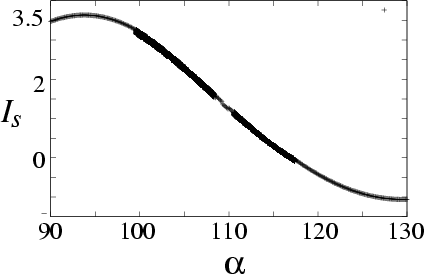}{\small (a)} \hfill
\includegraphics[width=0.45\columnwidth]{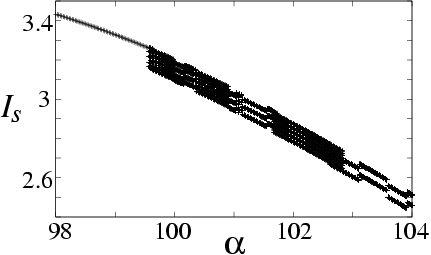}{\small (b)}
\caption{(a) The bifurcation diagram of the static VAR compensator
system. (b) Zoomed portion of the bifurcation diagram in the range
$\left(\alpha\in\left[98^{0},105^{0}\right]\right)$ showing the
occurrence of high-periodic orbits.}
\label{svar-bif}
\end{figure}

\begin{figure}[htb]
\centering \includegraphics[width=0.54\columnwidth]{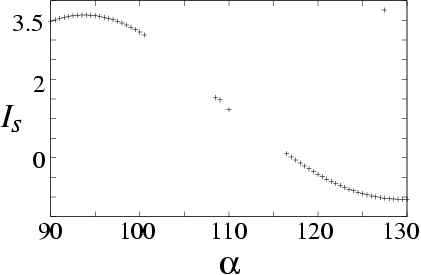}
\caption{The bifurcation diagram obtained by following the period-1
  orbit using the shooting method.}
\label{shooting}
\end{figure}

Let us now examine the effect of this discontinuity on the system's
bifurcation behavior.  A typical bifurcation diagram with $\alpha$ as
the variable parameter is presented in Fig.~\ref{svar-bif}(a). It is
seen that for $\alpha > 90^0$ and for $\alpha < 130^0$, the behavior
is period-1. But many peculiar dynamical transitions occur in the
intervening parameter range. In such a system it is possible to locate
the fixed point, irrespective of its stability, using a shooting
method \cite{donde-shooting-ijbc}. When the bifurcation diagram is
drawn by following the periodic orbit (Fig.~\ref{shooting}), it is
found that there is no fixed point in those parameter ranges. Even
though there is no fixed point, the orbit remains bounded and high
periodic orbits occur in a specific order (see the zoomed portion in
Fig.~\ref{svar-bif}(b)). The objective of this paper is to develop the
theory necessary to explain such atypical bifurcation behavior in
physical systems.

\section{The Normal Form}

Many hybrid dynamical systems  
can be represented in discrete-time by piecewise smooth maps,
given by equations of the form
\begin{eqnarray}
f(x,y,\mu) = \left\{\begin{array}{cc} g(x,y,\mu), & (x,y) \in R_A\\   
h(x,y,\mu), & (x,y) \in R_B \end{array} \right.
\label{eq:2dsys}
\end{eqnarray}
where $\mu$ is the bifurcation parameter and $R_A$ and $R_B$ are  
regions in the state space, divided by a borderline. We consider the
class of maps that have the following properties: 
\begin{enumerate}
\item The functions $g$ and $h$ are smooth (everywhere
differentiable),
\item The function $f$ is discontinuous across the borderline,
\item The elements of the Jacobian matrix of $f$ change discretely
across the borderline,
\item The Jacobian elements are finite.
\end{enumerate}
Maps of the above properties have application in many physical
and engineering systems (the system described in Section~\ref{var}
provides an example), and in the present paper we restrict our
attention to such maps.

The normal form for such a piecewise smooth
system in the neighborhood of a fixed point on the border can
be expressed as
\begin{equation}
G(x,y;\mu)=\left\{ \begin{array}{c}
\left(\begin{array}{cc}
\tau_{L} & 1\\
-\delta_{L} & 0\end{array}\right)\left(\begin{array}{c}
x\\
y\end{array}\right)+\mu\left(\begin{array}{c}
1\\
0\end{array}\right),\:\:\:\:\:\:\:\:\:\:\:\:\: x \leq 0\\
\left(\begin{array}{cc}
\tau_{R} & 1\\
-\delta_{R} & 0\end{array}\right)\left(\begin{array}{c}
x\\
y\end{array}\right)+(\mu+l)\left(\begin{array}{c}
1\\
0\end{array}\right),\:\;\:\:\: x>0\end{array}\right.
\label{eq:normal-map}
\end{equation}
where $\tau_L$, $\tau_R$ are the traces and $\delta_L$, $\delta_R$ are the
determinants of the Jacobian matrices at the two sides of the
border. The similar normal form of the piecewise smooth
continuous map was derived in \cite{nusse92,pre2d}. We have added the
discontinuity $l$ to obtain the normal form for the 2D discontinuous
map. We shall call the two halves of the state space $\{x<0\}$ and
$\{x>0\}$ as $L$ and $R$ respectively.

In order to explain the bifurcations in a specific system, one has to
obtain the eigenvalues of the fixed points at the two sides of the
borderline, and from that, the values of the trace and the
determinant. For example, in the case of the static VAR compensation
system described in Section~\ref{var}, the calculation of the Floquet
exponents at $\alpha = 100^\circ$ yielded $\tau_L=1.4677$,
$\delta_L=0.6550$, $\tau_R=1.4677$, and $\delta_R=0.6549$. The
explanation of the observed bifurcations have to be obtained by
analyzing the behavior of the normal form map for these parameter
values.

The fixed points of the system (\ref{eq:normal-map}) in both sides of
the boundary are given by:
\begin{eqnarray}
L^{*} & = & 
\left(\frac{\mu}{1+\delta_{L}-\tau_{L}},\frac{-\delta_{L}\mu}
{1+\delta_{L}-\tau_{L}}\right)\nonumber \\
R^{*} & = & 
\left(\frac{\mu+l}{1+\delta_{R}-\tau_{R}},\frac{-\delta_{R}(\mu+l)}
{1+\delta_{R}-\tau_{R}}\right)
\label{eq:fixed-pts}
\end{eqnarray}
If the $x$-component of $L^*$ is negative, the fixed point
exists. Else it does not. However, when the $x$-component of $L^*$ is
positive, iterations from initial conditions in the left half are
influenced by the ``nonexistent'' fixed point, which is called a
``virtual'' fixed point, and is denoted by $\bar{L}^*$. Similarly,
when the $x$-component of $R^*$ is positive, the fixed point exists;
else it is a virtual fixed point denoted by $\bar{R}^*$.

The stability of $L^*$ and $R^*$ are determined by the eigenvalues
$$\lambda_{L\pm}=
\frac{1}{2}\left(\tau_L\pm\sqrt{\tau_L^{2}-4\delta_L}\right), \;\;\;\;
\lambda_{R\pm}=\frac{1}{2}\left(\tau_R\pm\sqrt{\tau_R^{2}-4\delta_R}\right).$$

We shall confine our studies to dissipative systems, so that
$\left|\delta_{L}\right|<1$ and $\left|\delta_{R}\right|<1$.
Moreover, we shall be concentrating on positive determinant systems
because most physical systems are observed to yield maps with positive
determinant. For such a map, there can be four basic types of fixed
points:

\begin{enumerate}
\item When $-2\sqrt{\delta}<\tau<2\sqrt{\delta}$, both eigenvalues of
  the Jacobian are complex, with moduli less than $1$, indicating that
  the fixed point is a \textbf{spiral attractor}. If $\tau>0$ (real
  part positive), it is a clockwise spiral, and if $\tau<0$ (real part
  negative), the motion is counter-clockwise.
\item When $\delta<\tau^2/4$ and $-(1+\delta)<\tau<(1+\delta)$, both
eigenvalues are real and are less than $1$ in magnitude, causing the
fixed point to be an \textbf{attractor}. If
$-(1+\delta)<\tau<-2\sqrt{\delta}$, then we have $-1<\lambda_\pm<0$,
so that the attractor is a \textbf{flip attractor} (flips in
\emph{both} directions). If $2\sqrt{\delta}<\tau<(1+\delta)$, then
$0<\lambda_{+,-}<1$ and the fixed point is a \textbf{regular attractor}.
\item If $\tau<-(1+\delta)$, then $-1<\lambda_{+}<0$ and
$\lambda_{-}<-1$, so that the fixed point is a \textbf{flip saddle}
(flips in \emph{both} direction).
\item If $\tau>(1+\delta)$, then $\lambda_{+}>1$ and
$0<\lambda_{-}<1$, so that the fixed point is a \textbf{regular saddle}.
\end{enumerate}

Earlier work of PWS maps \cite{pre2d} demonstrated a property of the normal
form map that the unstable manifolds fold at every
intersection with the $x$-axis, and the image of every fold point is a
fold point. The stable manifolds fold at every intersection with the
$y$-axis and the pre-image of every fold point is a fold point
\cite{pre2d,cas2d}. In case of the discontinuous map, the same
property holds; the only difference is that the
folds are also discontinuous.

\section{Classification of discontinuous bifurcations}

We now classify the different types of discontinuous border collision
bifurcations depending on the type of fixed points occurring at the
two sides of the border. Unless otherwise stated, we shall
study the bifurcations occurring in the system (\ref{eq:normal-map})
as the parameter $\mu$ is varied. Note that while in a continuous map
($l=0$) border collision occurs at a single value of $\mu$, in the
discontinuous case ($l \neq 0$) border collision occurs at two values
$\mu=-l$ and $\mu=0$. Note also that the structure of the map is
different for positive and negative values of $l$. So in the
subsequent sections we shall consider the bifurcations for positive
and negative values of $l$ separately.

If a certain parameter combination exhibits a certain kind
of bifurcation behavior upon varying $\mu$ in one direction, then
interchanging the parameter values of the $L$ and $R$ sides will yield
the \emph{same} bifurcation behavior upon varying $\mu$ in the reverse
direction, i.e., interchanging parameter values is \emph{qualitatively}
mirror-symmetric with respect to the bifurcation behavior.

We now take up each case in the following subsections.

\subsection{Regular Saddle to Regular Saddle\protect \\
$\left(\tau_{L}>1+\delta_{L},\:\:\:\tau_{R}>1+\delta_{R}\right)$}

Since both the fixed points are unstable, in most situations the
 orbits starting from all initial conditions go to infinity. However,
 when the discontinuity is \emph{negative}, both $L^{*}$ and $R^{*}$
 exist in the range $\mu\in(0,-l)$. In such a situation, a trapping
 region can occur.

\begin{figure}[htb]
\includegraphics[width=0.45\columnwidth,height=0.35\columnwidth]
{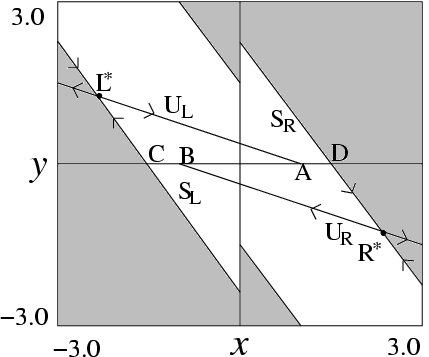}{\small (a)}\hfill
\includegraphics[width=0.45\columnwidth,height=0.35\columnwidth]
{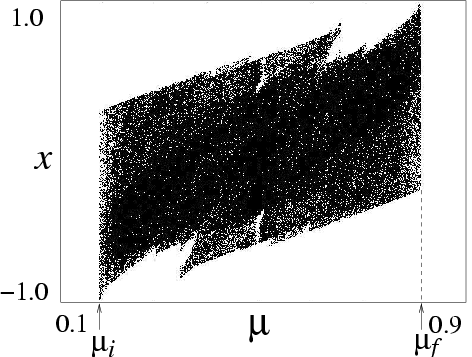}{\small (b)}
\caption{(a) Structure of the stable and unstable manifolds in the
  range $\mu\in(0,-l)$ when $L^*$ and $R^*$ are regular saddles. The
  parameter values are $\tau_{L}=\tau_{R}=1.5$, $\delta_{L}=\delta_{R}=0.3$,
  $l=-1$, $\mu=0.5$. The trapping region is shown in white. (b)
  Bifurcation Diagram showing the existence of only a chaotic orbit in
  the trapping region.
\label{fig:Basin1}}
\end{figure}

Let $U_{L}$ and $S_{L}$ be the unstable and stable manifolds of
 $L^{*}$ and $U_{R}$ and $S_{R}$ be the unstable and stable manifolds
 of $R^*$, respectively. The discontinuously folded structure of the
 manifolds can be seen in Fig.~\ref{fig:Basin1}(a).  Any initial state in
 $L$ to the \emph{right} of $S_{L}$ diverges away from $L^*$ along
 $U_{L}$ until it is mapped onto $R$. The state now comes under the
 influence of $R^*$ and is repelled back along $U_{R}$. The state
 therefore gets \emph{locked} between the two saddles, and this
 results in a bounded orbit. A similar behavior is seen for an initial
 state in $R$ to the \emph{left} of $S_{R}$. Since both sides are
 stretching, we observe a chaotic orbit. The chaotic attractor is
 stable as long as the unstable manifolds do not map to states outside
 the basin of attraction formed by $S_L$ and $S_R$. This happens in
 the parameter range $\left[\mu_{i},\mu_{f}\right]$ where
\begin{equation}
\mu_{i} =
\frac{\frac{-l\left(\lambda_{UR}-1\right)}{1-\tau_{R}+\delta_{R}}}
{\frac{1-\lambda_{SL}}{1-\tau_{L}+\delta_{L}}+\frac{\lambda_{UR}-1}
{1-\tau_{R}+\delta_{R}}},
\;\;\; 
\mu_{f} =
\frac{\frac{-l\left(1-\lambda_{SR}\right)}{1-\tau_{R}+\delta_{R}}}
{\frac{1-\lambda_{SR}}{1-\tau_{R}+\delta_{R}}+\frac{\lambda_{UL}-1}
{1-\tau_{L}+\delta_{L}}}\label{eq:bounds1}
\end{equation}
$\lambda_{UR}$ is the eigenvalue of $R^*$ outside the unit
circle, and $\lambda_{SR}$ is the eigenvalue of $R^*$ inside the unit
circle. $\lambda_{UL}$ and $\lambda_{SL}$ are the corresponding
eigenvalues of $L^*$. We will follow the same notations in the
subsequent sections. Thus $\mu_{i}$ and $\mu_{f}$ are the parameter
values at which \emph{boundary crisis} occurs, i.e., the unstable
manifolds $U_R$ and $U_{L}$ touch the stable manifolds $S_{L}$ and
$S_{R}$ respectively. We see in Fig.~\ref{fig:Basin1}(a) that the
unstable manifolds $U_{L}$ and $U_{R}$ meet the $x$-axis at $A$ and
$B$ respectively while the stable manifolds $S_{L}$ and $S_{R}$
intersect it at $C$ and $D$ respectively. Therefore $\mu_{i}$ is the
parameter value \emph{after} which $B$ lies towards the \emph{right}
of $C$ and $\mu_{f}$ is the value \emph{before} which $A$ lies towards
the $left$ of $D$, rendering the chaotic attractor stable. This range
of occurrence of the chaotic attractor is seen in the bifurcation
diagram of Fig.~\ref{fig:Basin1}(b). Note that if the map is
continuous, i.e., if $l=0$, the range of occurrence of the chaotic
orbit goes to zero, satisfying the situation described in
\cite{pre2d}.

\subsection{Attractor (Regular/Flip/Spiral) to Regular Saddle\protect \\
$\left(-\left(1+\delta_{L}\right)<\tau_{L}<\left(1+\delta_{L}\right),
\:\:\:\tau_{R}>1+\delta_{R}\right)$}

\noindent{\em The case of negative discontinuity:} 
For $\mu<0$, both $L^*$ and $R^*$ are real and we have a period-1
orbit. $L^{*}$ can be a regular/flip/spiral attractor depending on the
relationship between $\tau_{L}$ and $\delta_{L}$. The period-1
orbit is unique for $\mu<0$ if $L^*$ is a regular attractor. However,
high-periodic orbits as well as chaotic orbit may coexist with the
period-1 orbit for values of $\mu$ slightly less than $0$ if $L^*$ is
a spiral or flip attractor. When $L^{*}$ is very close to
the $y$-axis, any trajectory approaching it has a possibility of
going over to $R$ --- which happens for spiralling or flipping orbits.

\begin{figure}[htb]
\includegraphics[width=0.45\columnwidth,height=0.3\columnwidth]{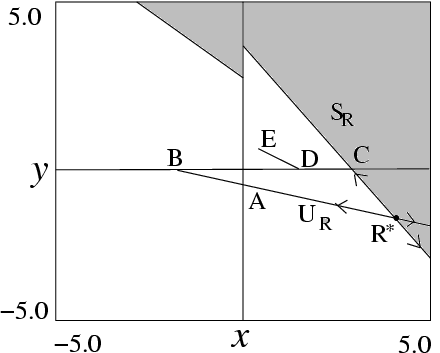}{\small (a)}\hfill
\includegraphics[width=0.45\columnwidth,height=0.3\columnwidth]{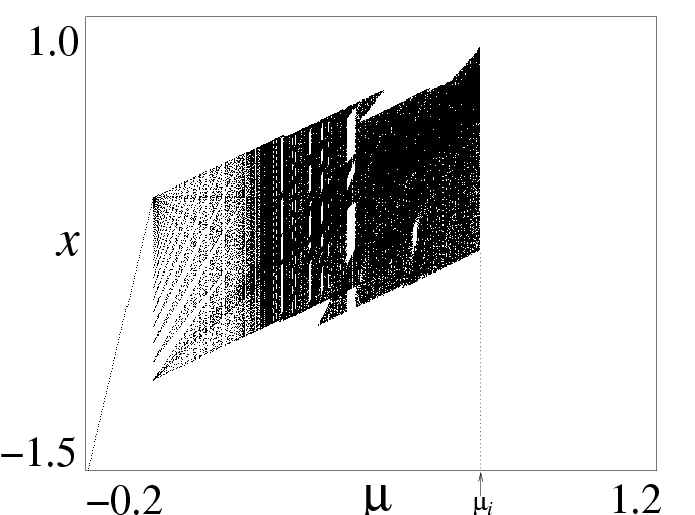}{\small (b)}
\caption{(a) The structure of stable and unstable manifolds when
  $m_{UR1}>m_{SR}$ and it is negative, $\tau_{L}=1.0,\:\tau_{R}=1.0,\:\delta_{L}=\delta_{R}=0.3,\:
l=-1,\:\mu=0.2$, and (b) Bifurcation diagram with
$\tau_{L}=1.0,\:\tau_{R}=1.5,\:\delta_{L}=0.1,\:\delta_{R}=0.2,\: l=-1$.
\label{fig:Basin2a}}
\end{figure}

\begin{figure}[htb]
\includegraphics[width=0.45\columnwidth,height=0.3\columnwidth]{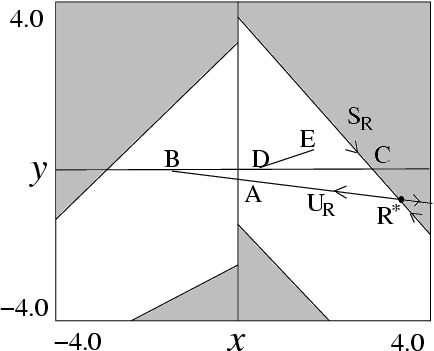}{\small (a)}\hfill
\includegraphics[width=0.45\columnwidth,height=0.3\columnwidth]{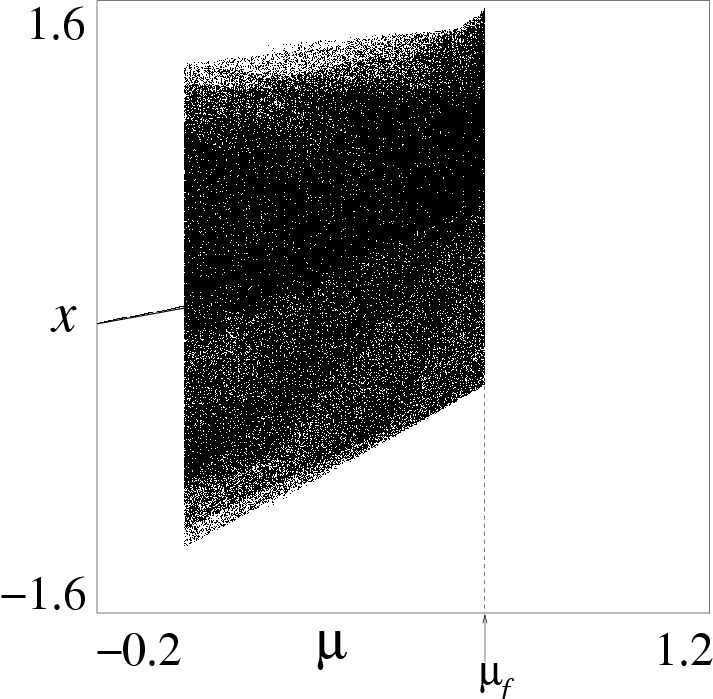}{\small (b)}
\caption{(a) The structure of stable and unstable manifolds when $m_{UR1}>m_{SR}$ but it is Positive,$\tau_{L}=-1.0,\:\tau_{R}=1.0,\:\delta_{L}=\delta_{R}=0.3,\:
l=-1,\:\mu=0.2$, and(b) Bifurcation diagram with
$\tau_{L}=-1.0,\:\tau_{R}=1.5,\:\delta_{L}=\delta_{R}=0.3,\: l=-1$.
\label{fig:Basin2b}}
\end{figure}

\begin{figure}[htb]
\includegraphics[width=0.45\columnwidth,height=0.3\columnwidth]{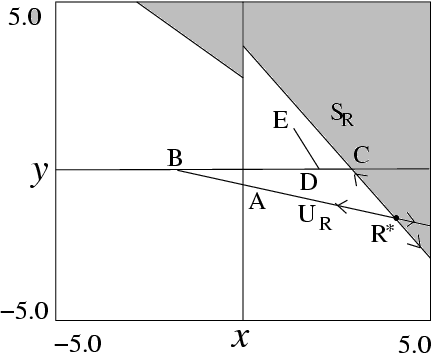}{\small (a)}\hfill
\includegraphics[width=0.45\columnwidth,height=0.3\columnwidth]{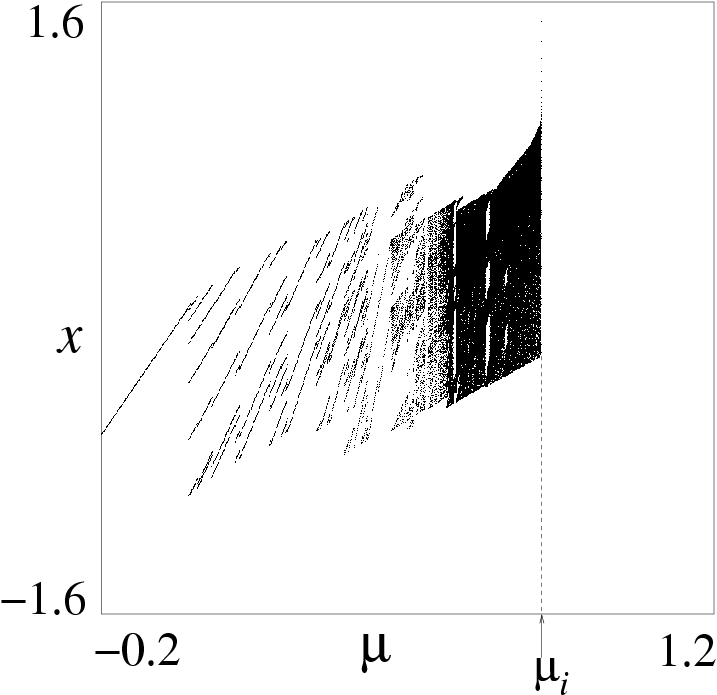}{\small (b)}
\caption{(a)The structure of stable and unstable manifolds when  $m_{UR1}<m_{SR}$,
$\tau_{L}=0.30,\:\tau_{R}=1.0,\:\delta_{L}=\delta_{R}=0.3,\:
l=-1,\:\mu=0.2$, and (b) Bifurcation diagram with
$\tau_{L}=0.30,\:\tau_{R}=1.5,\:\delta_{L}=\delta_{R}=0.3,\: l=-1$.
\label{fig:Basin2c}}
\end{figure}

For $\mu\in(0,-l)$, $\bar{L}^{*}$ is a virtual attractor, and $R^*$
is a regular saddle. The stable manifold $S_{R}$ of $R^{*}$ is
responsible for forming the basin boundary, while the virtual
attractor $\bar{L}^{*}$ present in $R$ draws the states in $L$ within
the basin of attraction towards $R$ (see Fig.~\ref{fig:Basin2a}(a),
Fig.~\ref{fig:Basin2b}(a), Fig.~\ref{fig:Basin2c}(a)). Any
initial state in $R$ to the \emph{left} of $S_{R}$ diverges
asymptotically along the unstable manifold $U_{R}$ until it is mapped
onto $L$. It now comes back under the influence of $\bar{L}^{*}$ and
is drawn back to $R$. The trajectory therefore gets locked between the
saddle and the virtual attractor resulting in a bounded orbit. This
orbit is stable before a parameter value $\mu_{c}$ after which the
unstable manifold $U_{R}$ maps to points outside the basin boundary
formed by $S_{R}$, i.e., a \emph{boundary crisis} occurs at
$\mu=\mu_{c}$ as $\mu$ is increased.
The value of $\mu_{c}$ depends on the slope of the unstable manifold
after it first folds discontinuously at the $x$-axis. This slope after
the first fold is
$m_{UR1}=\delta_{L}/\left(\lambda_{SR}-\tau_{L}\right)$ (Here
$\lambda_{SR}=\lambda_{R-}$). When $m_{UR1}>m_{SR}$, since both the
slopes are negative, the two manifolds can never intersect (see
Fig.~\ref{fig:Basin2a}(a)).

We see that $U_{R}$ intersects the $y$-axis at $A$, and arbitrarily
close points at the right and left of $A$ map discontinuously to $B$
and $D$ respectively. The attractor becomes unstable due to
boundary crisis when $D$ moves to the \emph{right} of $C$ on the
$x$-axis. This gives $\mu_{c}$ as

\begin{equation}
\mu_{c}=\mu_{i}=\frac{\left(2\lambda_{S{R}}-\delta_{R}-1\right)l}{\left(\lambda_{U{R}}-\lambda_{S{R}}\right)}\label{eq:case2ia}
\end{equation} 
which is positive.
However, under the conditions depicted in Fig.~\ref{fig:Basin2b}(a) and  Fig.~\ref{fig:Basin2c}(a), boundary crisis occurs when the segement of $U_{R}$  after the first fold \emph{intersects} with $S_{R}$. We see from the
corresponding Fig.~\ref{fig:Basin2b}(a) and  Fig.~\ref{fig:Basin2c}(a) that $U_{R}$ cuts the
$x$-axis at $B$, and $B$ maps to $E$. Thus the
bounded orbit becomes unstable when $E$ touches $S_{R}$. For this
condition, $\mu_{c}$ is given by
\begin{equation}
\mu_{c}=\mu_{f}=\frac{\left[\lambda_{U{R}}+\delta_{L}\left(1-\lambda_{U{R}}\right)+\tau_{L}\lambda_{U{R}}\left(1-\lambda_{U{R}}\right)-\lambda_{U{R}}\tau_{R}\right]l}
{\left[\delta_{R}\left(1+\lambda_{U{R}}\right)-\delta_{L}\left(1-\lambda_{U{R}}\right)+\tau_{L}\lambda_{U{R}}\left(1-\lambda_{U{R}}\right)-\lambda_{U{R}}\tau_{R}\right]}
\label{eq:case2ib}
\end{equation}


 As $\mu$ is increased through zero, the period-1 orbit vanishes through
 BCB and high periodic orbits of type $L^nR$ come into existence
 (Fig.~\ref{fig:Basin2a}(b) and Fig.~\ref{fig:Basin2c}(b)) with period
 increment and Farey tree sequence. With further increase of $\mu$
 symbol sequence gets reversed, i.e., the symbol sequence of type
 $LR^{n}$ come into existence with period adding and Farey tree. But
 since $\tau_{R} >(1+\delta_{R})$, eventually the orbit becomes
 globally stretching, leading to chaotic dynamics at some value of
 $\mu$. Depending on the value of $\tau_{L}$ a direct transition to
 chaos is also possible (Fig.~\ref{fig:Basin2b}(b)). In both the
 cases chaotic attractor disappears due to a boundary crisis.


\begin{figure}[htb]
\begin{center}
\includegraphics[width=0.55\columnwidth,height=0.4\columnwidth]{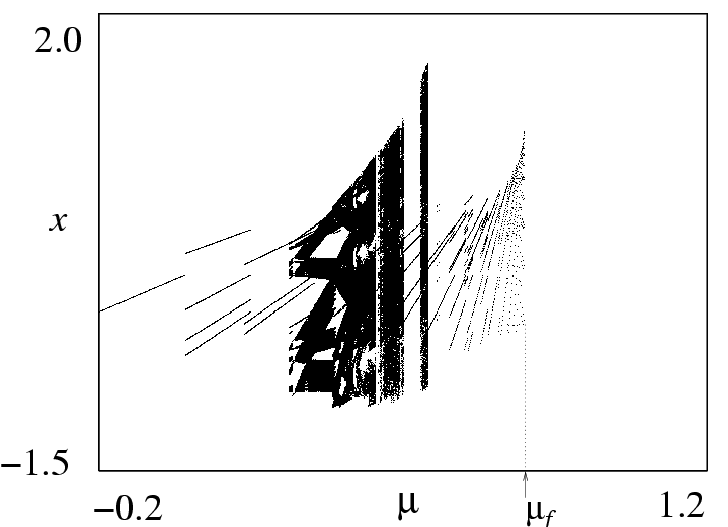}
\caption{ Bifurcation diagram with
$\tau_{L}=1.0,\:\tau_{R}=1.9,\:\delta_{L}=\delta_{R}=0.7,\: l=-1$.
\label{fig:bif1}}
\end{center}
\end{figure}
In the $\tau_L\!-\!\tau_R$ parameter space, the domains of occurrence
of each periodic orbit depends on the value of the determinants. For
higher values of the determinant the regions overlap, resulting in
coexistence of attractors and chaotic inclusions (see
Fig.~\ref{fig:bif1}).

 For $\mu>-l$, $L^*$ and $R^*$ are both virtual, which results in the
entire state space being unstable because initial conditions in $R$
diverge to infinity under the action of the unstable manifold
(directed away from $L$ to infinity) of the nonexistent fixed point
$\bar{R}^*$, now located inside $L$, and initial conditions in $L$ are
also attracted to the side $R$ due to the action of the nonexistent
virtual attractor $\bar{L}^*$.

For the case of positive discontinuity, we have a period-1 attractor
in the case when both $L^*$ and $R^*$ are real (for $\mu<-l$), else
the entire space is unstable.

The parameter space is symmetric upon variation of $\mu$ in the
reverse direction when $L^*$ is a regular saddle and $R^*$ is an
attractor, i.e., for $\tau_{L}>\left(1+\delta_{L}\right)$ and
$-\left(1+\delta_{R}\right)<\tau_{R}<\left(1+\delta_{R}\right)$. In
that case a chaotic attractor is initially born out of a boundary
crisis, slowly giving rise to increasingly stable orbits of
non-monotonically decreasing periodicities, ultimately leading to a
stable period-1 orbit.

\subsection{Regular Saddle to Flip Saddle\protect \\
$\left(\tau_{L}>1+\delta_{L},\:\:\:\tau_{R}<-\left(1+\delta_{R}\right)\right)$}

 The stable manifold $S_{L}$ of the regular saddle in $L$ is
responsible for forming the boundary of the basin of attraction, while
the flip saddle in $R$ is located inside this basin and is responsible
for creating the attractor. This leads to many interesting phenomena
as we shall see.

\begin{figure}[tbh]
\includegraphics[width=0.45\columnwidth,height=0.35\columnwidth]
{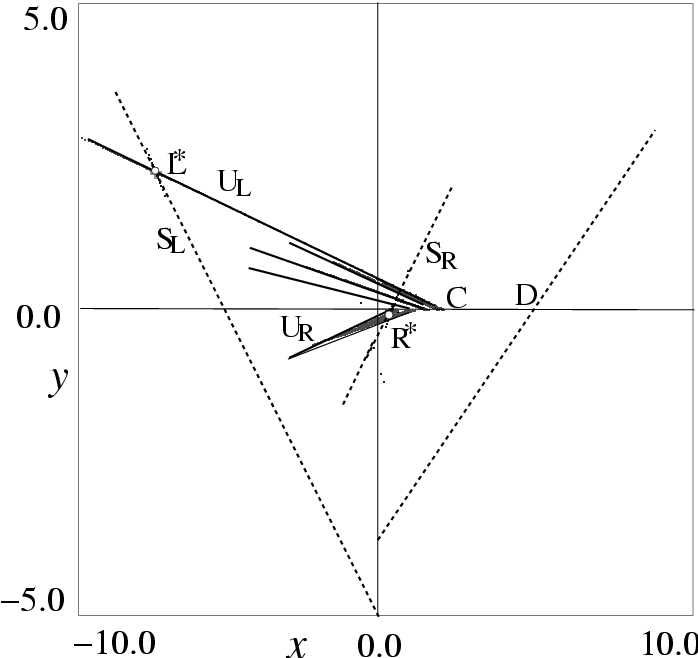}{\small (a)}\hfill
\includegraphics[width=0.45\columnwidth,height=0.35\columnwidth]
{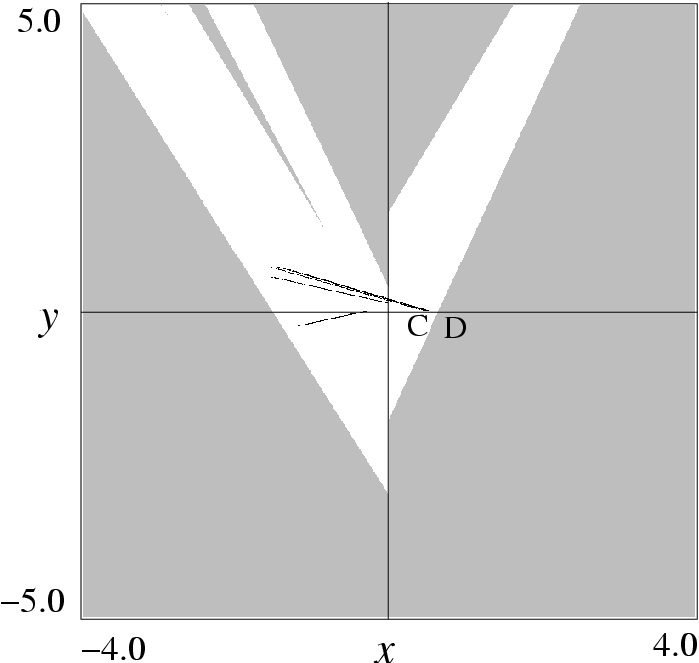}{\small (b)}
\caption{(a) Homoclinic and heteroclinic intersections of the stable
and unstable manifolds of the flip saddle $R^*$ when
$\tau_{L}=1.5$,\:$\tau_{R}=-1.5$,\:$\delta_{L}=\delta_{R}=0.3$,\:
$l=-1$,\:$\mu=1.5$. This figure and the subsequent ones showing the
invariant manifolds have been generated using DsTool \cite{back1,osinga2,osinga1}. (b) The chaotic attractor, and the basin of
attraction created by the stable manifold of $L^*$, for
$\tau_{L}=1.5$,\:$\tau_{R}=-1.5$,\:$\delta_{L}=\delta_{R}=0.3$,\:
$l=-1$,\:$\mu=0.5$.
\label{fig:Homocl3}}
\end{figure}

 {\em The case of negative discontinuity:} 
For $\mu<0$, when
both saddles are virtual, we do not observe any attractor. For
$\mu>-l$, $L^{*}$ is a regular saddle and $R^{*}$ is a flip
saddle. All initial conditions in $L$ converge onto the unstable
manifold $U_{L}$. An initial condition to the \emph{right} of the
basin boundary formed by $S_{L}$ converges onto the unstable manifold
$U_{L}$ and is mapped into $R$ (see Fig. \ref{fig:Homocl3}(a)). The
slope of the unstable manifold of $R^{*}$ is
$m_{U{R}}=-\delta_{R}/\lambda_{U{R}}=-\lambda_{R+}$ while the slope of
the stable manifold is $m_{S{R}}=-\lambda_{R-}$. The unstable manifold
$U_{R}$ folds discontinuously at the intersection with the $x$-axis
and continues with a slope
$m_{U{R}}=\delta_{L}\lambda_{U_{R}}/\left(\delta_{R}-\tau_{L}
\lambda_{U{R}}\right)$
$=\delta_{L}\lambda_{R-}/\left(\delta_{R}-\tau_{L}\lambda_{R-}\right)$.
Since $m_{S{R}}>0$ and $m_{U{R}}<0$ (as dictated by the range of
parameters in question), the unstable manifold $U_R$ must have a
transverse homoclinic intersection with the stable manifold $S_R$
after the first fold, which implies an infinity of homoclinic
intersections and the existence of a chaotic orbit. For the regular
saddle $L^{*}$, we note that the unstable manifold $U_{L}$ has a
negative slope $m_{U{L}}=-\lambda_{L+}$, and must, therefore, also
have a heteroclinic intersection with the stable manifold $S_{R}$ of
the flip saddle $R^{*}$. The Lambda Lemma \cite{yorkebook} implies
that the unstable manifolds of the two sides will come arbitrarily
close to each other, which makes
the chaotic attractor robust \cite{robust}.

It is clear from the geometric structure that no point of the
attractor can be to the right of point $C$. If $C$ lies towards the
left of $D$, the chaotic orbit is stable. If $C$ falls outside the
basin of attraction, it is an \emph{unstable chaotic orbit}. From this we obtain the critical value of $\mu$
for which boundary crisis will occur:
\begin{equation}
\mu_{i}=\frac{-\lambda_{U_{L}}\left(1-\tau_{L}+\delta_{L}\right)l}
{\lambda_{U_{L}}\left(\lambda_{S_{L}}-\tau_{L}+\delta_{L}\right)
+\left(\lambda_{U_{L}}-1\right)\left(\delta_{R}-\lambda_{U_{L}}\tau_{R}\right)}
\label{eq:stabchaotic}
\end{equation}
which is the same as (\ref{eq:case2ib}) taken in the reverse direction of $\mu$
variation. Note that for a continuous map the occurrence
of the boundary crisis does not depend on $\mu$, but in the discontinuous
map it does.

The same phenomenon occurs for $0<\mu<-l$ where $\bar{R}^{*}$ is a
virtual flip saddle, but its stable and unstable manifolds still exist
in $R$ and undergo homoclinic intersections resulting in chaos. The
chaotic orbit 
occurs inside the basin of attraction formed by the stable manifold of
$L^*$ (see Fig. \ref{fig:Homocl3}(b)).

{\em The case of positive discontinuity:} 
For $\mu>0$, $L^{*}$ is a
regular saddle and $R^{*}$ is a flip saddle. In this case the same
mechanism as discussed above causes a chaotic attractor, with the
same stability condition as in (\ref{eq:stabchaotic}).

When there is a transition from a flip saddle to a regular saddle,
i.e., when $\tau_{L}<-\left(1+\delta_{L}\right)$ and
$\tau_{R}>1+\delta_{R}$, the bifurcation behavior is the same if $\mu$
is varied in the opposite direction.

\subsection{Attractor to Attractor (Regular/Flip/Spiral)\protect \\
$-\left(1+\delta_{L}\right)<\tau_{L}<\left(1+\delta_{L}\right),\:\:\:
-\left(1+\delta_{R}\right)<\tau_{R}<\left(1+\delta_{R}\right)$
\label{att-to-att}}

Once again, in this case, we shall be led to observe 
phenomena that do not occur in smooth maps or piecewise smooth
continuous maps. 

{\em The case of negative discontinuity:} For both $\mu<0$ and
$\mu>-l$, when one of the attractors is real and the other virtual, a
stable period-1 orbit exists. If the fixed point is a spiral or flip
attractor, high-periodic orbits may coexist with the period-1 orbit
for values of $\mu$ slightly less than zero or slightly greater than
$-l$, because initial conditions close to the border may map to the
other side before converging on the fixed point.

\begin{figure}[tbh]
\includegraphics[width=0.45\columnwidth,height=0.35\columnwidth]{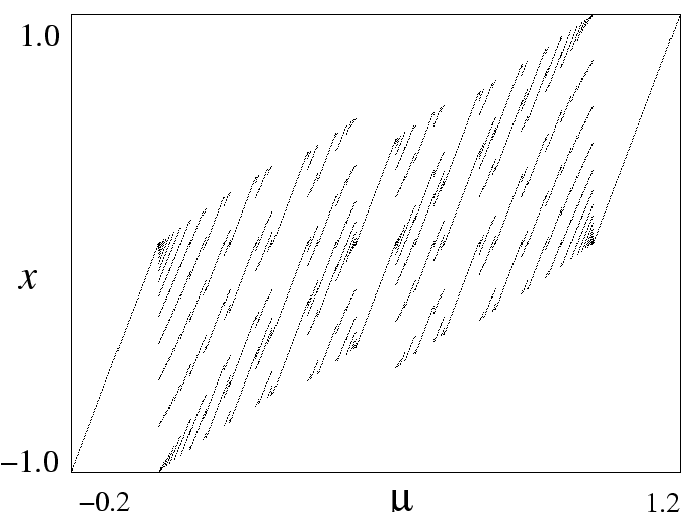}{\small (a)}\hfill
\includegraphics[width=0.45\columnwidth,height=0.35\columnwidth]{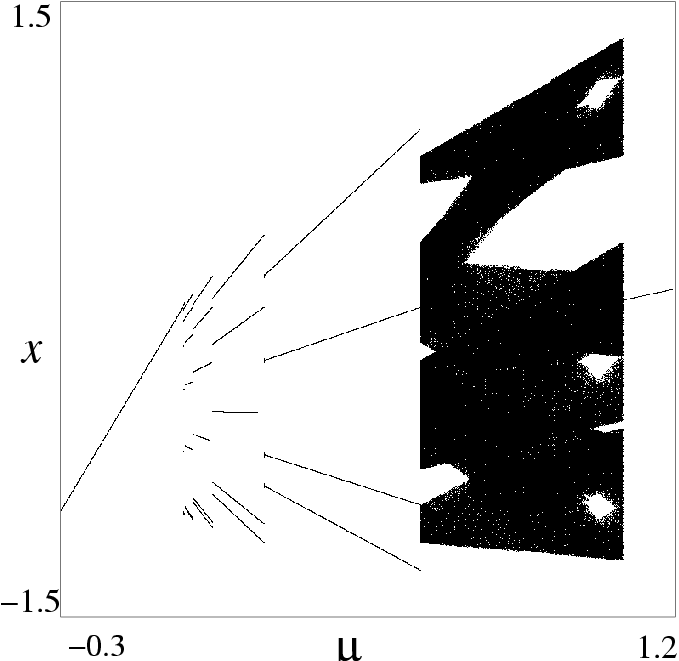}{\small (b)}
\caption{(a) Bifurcation Diagram when both $\tau_{L}$\: and\:
$\tau_{R}$ are positive, with $\tau_L=1.0$, $\delta_L=0.2$,
$\tau_R=1.$, $\delta_R=0.2$, and $l=-1$. (b) Bifurcation Diagram when
$\tau_{L}$ is positive and $\tau_{R}$ is negative, with
$\tau_{L}=1.0$, $\tau_{R}=-1.0$, $\delta_{L}=\delta_{R}=0.3$, $l=-1$.
\label{fig:Bif-4a}}
\end{figure}

For $\mu\in(0,-l)$, both $\bar{L}^{*}$ and $\bar{R}^{*}$ are virtual
attractors. Thus any initial state in $L$ is drawn towards the virtual
attractor $\bar{L}^{*}$ situated in $R$. However as soon as it crosses
the $y$-axis, it is drawn back towards the virtual attractor
$\bar{R}^{*}$ situated in $L$. The state therefore gets \emph{locked}
between the two virtual attractors and that results in a bounded
orbit. The same behavior is exhibited by an initial state in $R$. This
closed orbit may be high-periodic or chaotic.  For
$0<\tau_{L}<\left(1+\delta_{L}\right)$ and
$0<\tau_{R}<\left(1+\delta_{R}\right)$, the bifurcation diagram in
Fig.~\ref{fig:Bif-4a}(a) shows that as $\mu$ is increased through zero
period-1 orbit vanishes through border collision bifurcation and high
periodic orbits of type $L^nR$ come into existence with period
inclusion and Farey tree sequence. With further increase of $\mu$, the
symbol sequence gets reversed, i.e., the symbol sequence of type
$LR^n$ comes into existence. Finally a period-1 orbit come into
existence through BCB.

\begin{figure}[tbh]
\includegraphics[width=0.45\columnwidth,height=0.35\columnwidth]{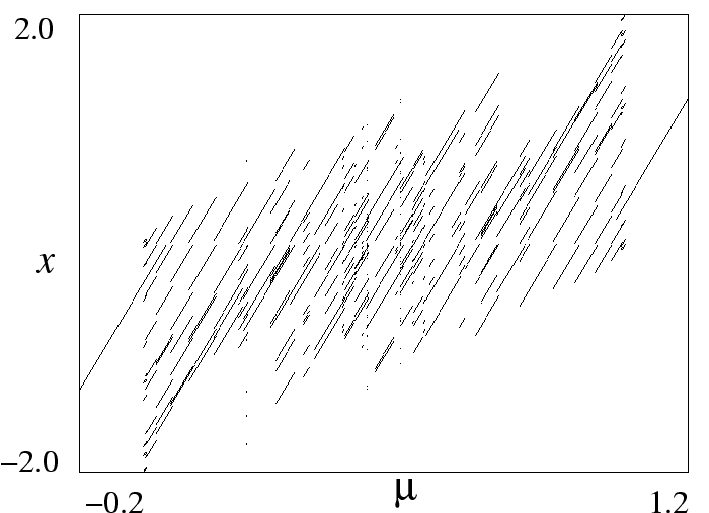}{\small (a)}\hfill
\includegraphics[width=0.45\columnwidth,height=0.35\columnwidth]{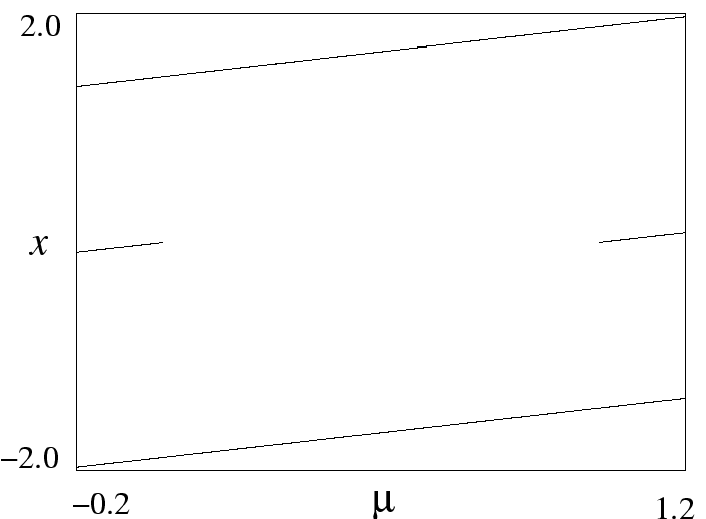}{\small (b)}
\caption{(a) Bifurcation Diagram when both $\tau_{L}$\: and\:
$\tau_{R}$ are positive, with $\tau_{L}=\tau_{R}=1.4677$,
$\delta_{L}=\delta_{R}=0.655$, $l=-1$. (b) Bifurcation Diagram when
both $\tau_{L}$\: and\: $\tau_{R}$ are negative, with
$\tau_{L}=\tau_{R}=-1.0$, $\delta_{L}=\delta_{R}=0.3$, $l=-1$.
\label{fig:Bif-4b}}
\end{figure}

We have seen that in a discontinuous map a parameter range can exist
where both the fixed points are virtual, and if both the virtual fixed
points are attracting in nature, high periodic orbits occur in
specific sequence. 

However for higher values of the determinants
($\delta_{L},\delta_{R}$), the bifurcation structure is more complex. Here
we can observe period increment with coexistence of attractors. Notice
that in Fig.~\ref{fig:Bif-4b}(a)  the parameters are the same as the
ones obtained for the physical system considered in Section~2. The
traces and determinants obtained from the physical system indicate
that the local behavior of the system satisfies the condition being
discussed in this section. The dynamical phenomena observed in
Fig.~\ref{svar-bif} and in Fig.~\ref{fig:Bif-4b}(a) are qualitatively
the same, which explains the peculiar dynamical behaviors observed in
the static VAR controller system.

\begin{figure}[tbh]
\includegraphics[width=0.45\columnwidth,height=0.35\columnwidth]{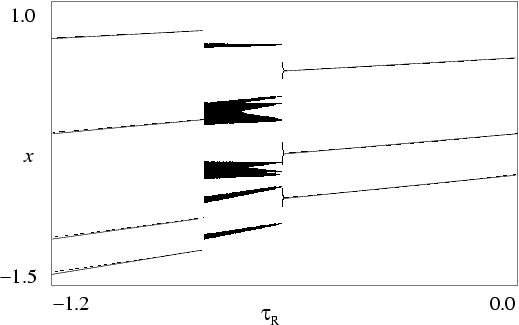}{\small (a)}\hfill
\includegraphics[width=0.45\columnwidth,height=0.35\columnwidth]{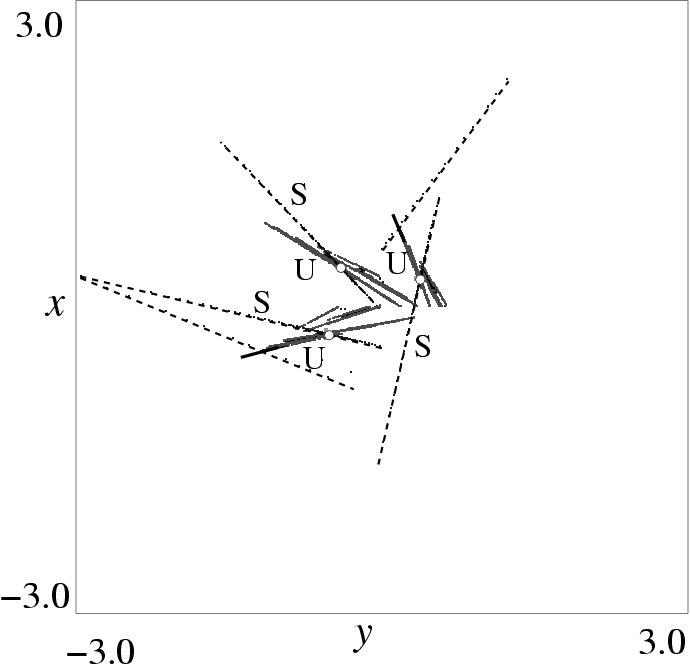}{\small (b)}
\caption{(a) Bifurcation diagram with respect to $\tau_{R}$, for $\tau_{L}=1.0$, $\delta_{L}=\delta_{R}=0.3$, $l=-1$, $\mu=0.5$. (b) Homoclinic
intersection of stable and unstable manifolds of a period-3 saddle.
\label{fig:Bif4t-hom}}
\end{figure}

For $0<\tau_{L}<\left(1+\delta_{L}\right)$ and
$-\left(1+\delta_{R}\right)<\tau_{R}<0$, the high period orbits do not
occur throughout the parameter range and a chaotic orbit is found to
occur (see the bifurcation diagram in Fig.~\ref{fig:Bif-4a}(b)) for
high values of $\mu$. This is a novel phenomenon --- the occurrence of
chaos in a system where both parts of the state space are contractive.
The mechanism of the creation of the chaotic orbit becomes clear from
the bifurcation diagram with $\tau_R$ as the parameter (see
Fig.~\ref{fig:Bif4t-hom}(a)). As the parameter is reduced, the
period-3 orbit's eigenvalues reach $-1$ and it undergoes a period
doubling. The resulting period-6 orbit hits the border and turns into
a six-piece chaotic orbit. The period-3 fixed point is now a saddle,
and Fig.~\ref{fig:Bif4t-hom}(b) shows that its stable and unstable
manifolds undergo a homoclinic intersection. This gives rise to the
chaotic orbit. At a lower value of the parameter, a period-4 orbit
starts to exist, initially coexisting with the chaotic orbit. At a
specific parameter value, a boundary crisis occurs and the chaotic
attractor disappears. At other parameter ranges, a similar phenomenon
occurs for the other high-periodic fixed points.

For $-\left(1+\delta_{L}\right)<\tau_{L}<0$ and
$-\left(1+\delta_{R}\right)<\tau_{R}<0$, i.e., when both the fixed
points are flip attractors, a different situation occurs. In the
parameter range $\mu\in(0,-l)$ when no fixed point exists (both are
virtual), only a period-2 orbit occurs (see
Fig.~\ref{fig:Bif-4b}(b)). In fact, the range of
occurrence of the period-2 orbit extends beyond this range, and
the period-2 orbit coexists with the period-1 orbit over a significant
parameter range. Note that in the case of a continuous map, a stable
period-2 orbit cannot occur when the map is globally contractive
\cite{feigin-mario,pre2d}.

A symmetric behavior is exhibited for
$-\left(1+\delta_{R}\right)<\tau_{R}<0$ and
$0<\tau_{L}<\left(1+\delta_{L}\right)$ upon varying $\mu$ in the
reverse direction.

\begin{figure}[tbh]
\includegraphics[width=0.45\columnwidth,height=0.35\columnwidth]{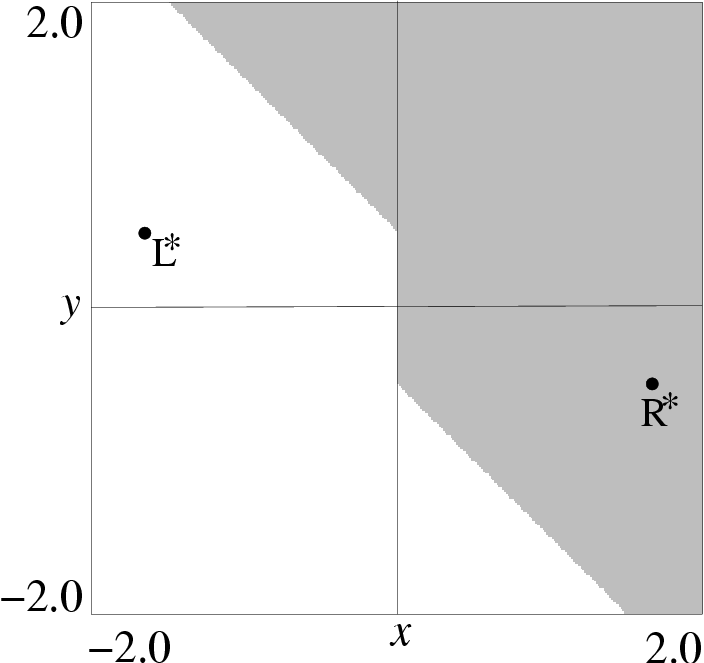}{\small (a)}\hfill
\includegraphics[width=0.45\columnwidth,height=0.35\columnwidth]{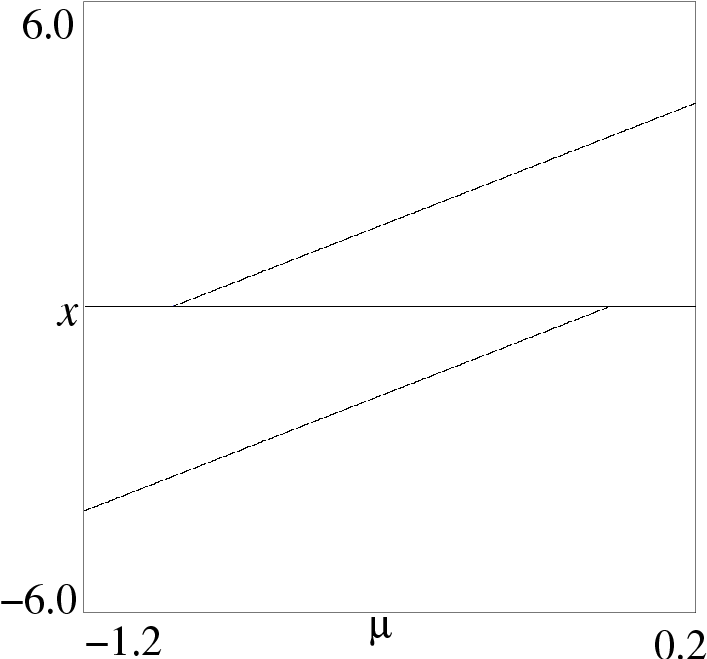}{\small (b)}
\caption{(a) Basin of attraction, for $\tau_{L}=1.0$, $\tau_{R}=1.0$, $\delta_{L}=\delta_{R}=0.3$, $\mu=0.5$, $l=1.0$. (b) Bifurcation diagram with respect to $\mu$, for $\tau_{L}=1.0$, $\tau_{R}=1.0$, $\delta_{L}=\delta_{R}=0.3$, $l=1.0$.
\label{fig:Basin4}}
\end{figure}

{\em The case of positive discontinuity:} In this case there is one
real attracting fixed point in each of the parameter ranges $\mu<-l$
and $\mu>0$.  For $\mu\in(0,-l)$ both ${L}^{*}$ and ${R}^{*}$ are
attracting fixed points, and have their own basins of attraction. It
is known that the basin boundary is normally formed by the stable
manifold of a saddle fixed point. However, under the condition
$0<\tau_{L}<(1+\delta_L)$ and $0<\tau_{R}<(1+\delta_R)$ the basin
boundary is formed by a completely different mechanism. In this case
the segment of the borderline from $-(\mu+l)$ and $-\mu$ (the two
preiterates of the origin) form a part of the basin boundary. The rest
of the basin boundary is formed by successive preiterates of this
segment, as can be seen in Fig.~\ref{fig:Basin4}(a). This is due to
the fact that in a discontinuous map, the borderline can act as a
repellor. Moreover, its preiterates can also act as repellors since
two points at the two sides of these line segments eventually map to
points far apart.  This mechanism gives rise to the bifurcation
diagram of Fig.~\ref{fig:Basin4}(b).

\subsection{Regular Attractor to Flip Saddle\protect \\
$\left(2\sqrt{\delta_{L}}<\tau_{L}<\left(1+\delta_{L}\right),\:\:\:
\tau_{R}<-\left(1+\delta_{R}\right)\right)$}

{\em The Case of Negative Discontinuity:}
 For $\mu<0$, $L^{*}$ is a regular attractor and $\bar{R}^{*}$ is a
virtual flip saddle, and the attracting fixed point enforces a stable period-1
orbit since all initial conditions eventually lead to it.

\begin{figure}[tbh]
\includegraphics[width=0.45\columnwidth,height=0.35\columnwidth]{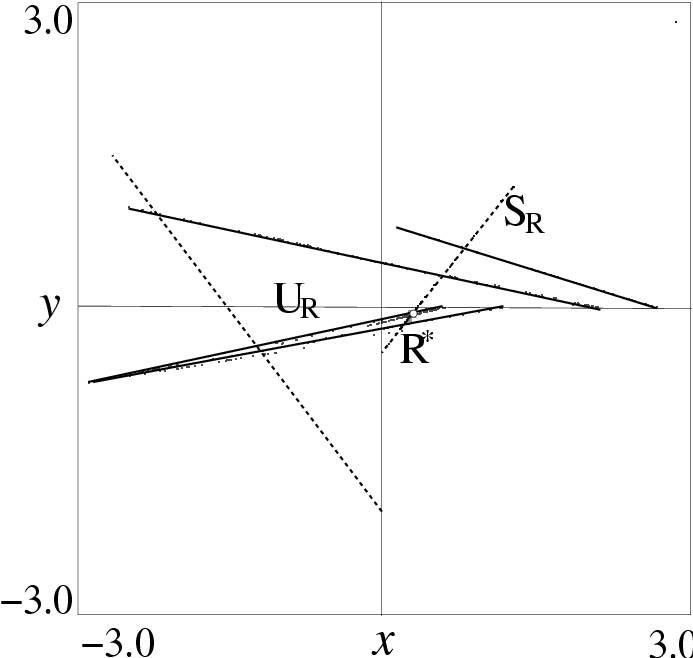}{\small (a)}\hfill
\includegraphics[width=0.45\columnwidth,height=0.35\columnwidth]{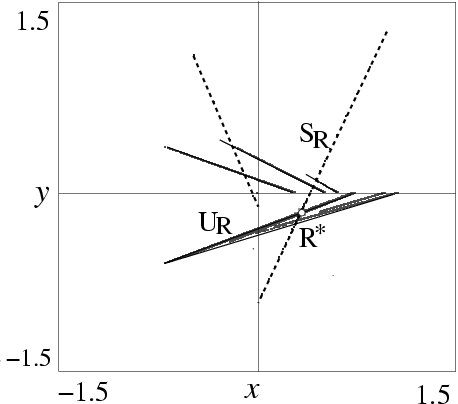}{\small (b)}
\caption{(a) Homoclinic intersections of stable and unstable manifolds
of the flip saddle $R^*$ for a negative discontinuity, with
$\tau_{L}=1.0,\:\tau_{R}=-1.0,\:\delta_{L}=\delta_{R}=0.3,\:l=-1,\:\mu=0.5$.
(b) Homoclinic intersections of stable and unstable manifolds of the
flip saddle $R^*$ for a positive discontinuity, with
$\tau_{L}=1.2,\:\tau_{R}=-1.5,\:\delta_{L}=\delta_{R}=0.3,\:l=1,\:
\mu=0.5$.  
\label{fig:Homocl5}}
\end{figure}

For $\mu>-l$, $\bar{L}^{*}$ is a virtual attractor, and hence lies in
$R$. $R^{*}$ on the other hand is a flip saddle, and its stable and
unstable manifolds undergo homoclinic intersections after the first
fold (see Fig.~\ref{fig:Homocl5}(a)) to yield a chaotic attractor.
After the first fold of $U_{R}$, its slope becomes
$\delta_{L}\lambda_{U_{R}}/\left(\delta_{R}-\tau_{L}\lambda_{U_{R}}\right)$
(a negative value), while the slope of $S_{R}$ given by
$-\lambda_{U_{R}}$ remains positive. All initial conditions in $L$ are
drawn to $R$ due to the action of the virtual attractor $\bar{L}^{*}$,
where they converge onto the chaotic attractor. This chaotic attractor
extends into the parameter range $0<\mu<-l$, when $\bar{R}^{*}$
becomes virtual, but its stable and unstable manifolds still exist in
$R$ and undergo homoclinic intersections to yield a chaotic
orbit. Hence, any initial condition in the entire phase space
ultimately converges on the chaotic attractor. The chaotic attractor
manifests for most of the parameter space, but for values of $\mu$
slightly greater than zero some high-periodic orbits are also stable.

{\em The Case of Positive Discontinuity:}
For $\mu<-l$, $L^{*}$ is a regular attractor and $\bar{R}^{*}$ is a
virtual flip saddle; for $\mu\in(-l,0)$, $L^{*}$ is a regular
attractor and $R^{*}$ is a flip saddle; for $\mu>0$, $\bar{L}^{*}$ is
a virtual regular attractor and $R^{*}$ is a flip saddle.
For the first two cases, the regular attractor in $L$ causes a
period-1 orbit to exist for all initial conditions. For the case when
$\bar{L}^{*}$ is a virtual attractor and $R^{*}$ is a flip saddle, the
stable and unstable manifolds of $R^{*}$ undergo homoclinic intersection
(Fig.~\ref{fig:Homocl5}(b)) to yield a chaotic orbit.

\subsection{Spiral Attractor to Flip Saddle\protect \\
$\left(-2\sqrt{\delta_{L}}<\tau_{L}<2\sqrt{\delta_{L}},\:\:\:
\tau_{R}<-\left(1+\delta_{R}\right)\right)$}

\begin{figure}[tbh]
\includegraphics[width=0.45\columnwidth,height=0.35\columnwidth]
{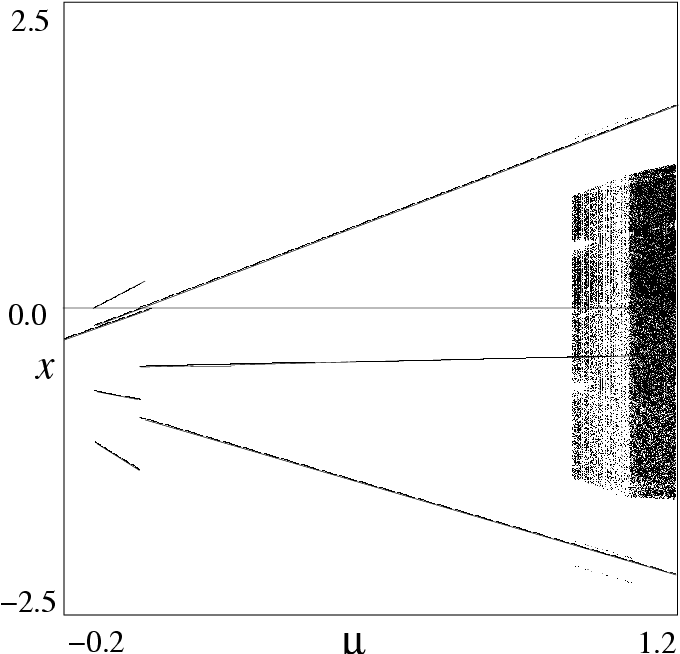}{\small (a)}\hfill
\includegraphics[width=0.45\columnwidth,height=0.35\columnwidth]
{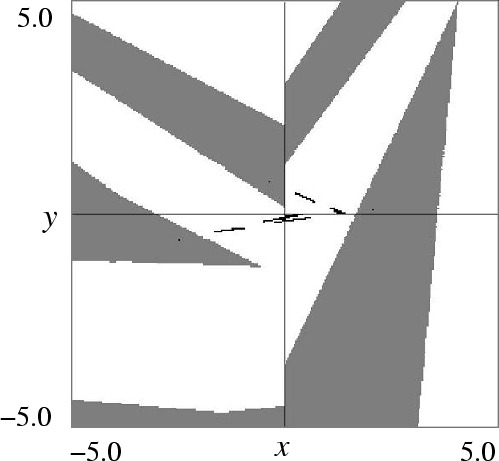}{\small (b)}
\caption{(a) Bifurcation Diagram with
$\tau_{L}=0.5$,\:$\tau_{R}=-1.5$,\:$\delta_{L}=\delta_{R}=0.3$,\:$l=-1$. 
(b) Basin of Attraction with
$\tau_{L}=0.5$,\:$\tau_{R}=-1.5$,\:$\delta_{L}=\delta_{R}=0.3$,\:
$l=-1$,\:$\mu=1.2$. 
\label{fig:case6}}
\end{figure}

{\em The Case of Negative Discontinuity:}
  For $\mu<0$, $L^*$ is a spiral attractor and $\bar{R}^*$ a virtual
flip saddle, and the attractor causes a period-1 orbit. Whenever the
spiral attractor is very close to the $y$-axis (which happens either
when $\mu$ slightly less than zero or $\tau_{L}$ is negative),
high-periodic orbits coexist with the period-1 orbit (see
Fig.~\ref{fig:case6}(a)).

High-periodic orbits continue to exist for $\mu\in(0,-l)$ when
$\bar{L}^{*}$ is a virtual attractor and $\bar{R}^{*}$ is a virtual
flip saddle, because initial conditions in $L$ are mapped to $R$ due
to the virtual attractor $\bar{L}^{*}$ in $R$, and points in $R$ are
then mapped back to $L$ due to action of the virtual flip saddle
$\bar{R}^{*}$, thus leading to a bounded orbit, which manifests as a
high-periodic or chaotic orbit (chaos occurs only when $\tau_{L}$ is
positive).

For $\mu>-l$, $\bar{L}^{*}$ is a virtual attractor while $R^{*}$ is a
flip saddle. The virtual attractor $\bar{L}^{*}$ maps all initial
conditions in $L$ to $R$, which subsequently flip to $L$. This may
either lead to a stable high-periodic orbit or a chaotic orbit. The basin of
attraction in Fig.~\ref{fig:case6}(b) shows that there are high
periodic attractors coexisting with a chaotic attractor when
$\mu>-l$. It is found that in much of the parameter range, the
conditions for the occurrence of period-2 or period-3 are satisfied.
The only difference between the cases when $\tau_{L}$ is positive
$\left(0<\tau_{L}<2\sqrt{\delta_{L}}\right)$ and when it is negative
$\left(-2\sqrt{\delta_{L}}<\tau_{L}<0\right)$ (which basically implies a
change in the sense of rotation of the spiral attractor $L^{*}$) is
that in the negative case, a period-2 orbit is found to satisfy the
conditions for stability and existence for the entire range of $\mu$.

{\em The Case of Positive Discontinuity:}
 For $\mu<-l$, $L^{*}$ is a spiral attractor while $\bar{R}^{*}$ is a
virtual flip saddle; for $\mu\in(-l,0)$, $L^{*}$ is a spiral attractor
and $R^{*}$ is a flip saddle; for $\mu>0$, $\bar{L}^{*}$ is a virtual
attractor while $R^{*}$ is a flip saddle. The first two cases will
cause a stable period-1 orbit to exist. For the last case, the virtual
attractor $\bar{L}^{*}$ maps all initial conditions in $L$ to $R$,
which subsequently flip to $L$. This may either lead to a stable
high-periodic orbit or a chaotic orbit.

\subsection{Flip Attractor to Flip Saddle\protect \\
$\left(-\left(1+\delta_{L}\right)<\tau_{L}<-2\sqrt{\delta_{L}},\:\:\:\tau_{R}<-\left(1+\delta_{R}\right)\right)$}

{\em The Case of Negative Discontinuity:}
For $\mu<0$, $L^{*}$ is a flip attractor while $\bar{R}^{*}$ is a
virtual flip saddle; for $\mu\in(0,-l)$, $\bar{L}^{*}$ is a virtual
flip attractor and $\bar{R}^{*}$ is a virtual flip saddle; for
$\mu>-l$, $\bar{L}^{*}$ is a virtual flip attractor and $R^{*}$ is a
flip saddle. For the first case a stable period-1 orbit exists. For
the other two cases, all initial conditions in $L$ flip to the other
side of $\bar{L}^{*}$ and land in $R$. Points in $R$ diverge to
infinity along the unstable manifold $U_{R}$ of the flip saddle (no
matter whether it is real or virtual), so that no attractor exists in
the phase space.

{\em The Case of Positive Discontinuity:} For $\mu<-l$, $L^{*}$ is a
 flip attractor and $\bar{R}^{*}$ is a virtual flip saddle. A stable
 period-1 orbit exists but higher periodic orbit may coexist along
 with the period-1 orbit.  For $\mu\in(-l,0)$, $L^{*}$ is a flip
 attractor and $R^{*}$ is a flip saddle, and a period-1 orbit exists
 and almost the entire space is stable.  For $\mu>0$, $\bar{L}^{*}$ is
 a virtual flip attractor while $R^{*}$ is a flip saddle, and no
 attractor exists.  There is no basin of attraction
 because of the diverge-to-infinity action of the unstable manifold of
 the flip saddle $R^{*}$ similar to the cases discussed earlier (since
 the attractor is virtual).

The bifurcation behavior for the case
$\tau_{L}<-\left(1+\delta_{L}\right)$ and
$-\left(1+\delta_{R}\right)<\tau_{R}<-2\sqrt{\delta_{R}}$  is symmetric
upon varying the parameter $\mu$ in the opposite direction.

\subsection{Flip Saddle to Flip Saddle (flipping along \emph{both} directions)\protect \\
$\left(\tau_{L}<-\left(1+\delta_{L}\right),\:\:\:\tau_{R}<-\left(1+\delta_{LR}\right)\right)$}

No attractor exists in this case for any parameter value for both
negative and positive discontinuity because of the action of the
unstable manifolds of both the saddles---real or virtual.

\section{Conclusion}

This paper contains the first attempt to analyze the bifurcation
phenomena in two-dimensional discontinuous maps in terms of
asymptotically stable behavior occurring for various parameter
combinations. In this investigation we have used a
piecewise affine approximation of the map in the neighborhood of the
border, and then have partitioned the parameter space into regions
depending on the types of the fixed points at the two sides of the
border. For each case we have described the ``typical'' bifurcation
phenomena in terms of the stable orbits. Where applicable, we have
explained the occurrence and stability of chaotic orbits in terms of
the structure of the stable and unstable manifolds of the fixed
points. We do not claim to have done a complete study. Indeed,
the dynamics of this system is so rich that it may take years to
investigate all the possible situations. 

A few atypical features of the bifurcations in discontinuous maps have
been found in this study:
\begin{enumerate}
\item The bifurcation behavior depends on the sign of the
  discontinuity;
\item There can be stable orbits even when no fixed point exists;
\item Chaos can occur even when each subsystem is contractive in
  nature;
\item Period-incrementing sequences are common. Each transition is
  ``hard'' in the sense that the state discontinuously jumps from one
  periodic orbit to another.
\item While in continuous maps, only stable manifolds of saddle fixed
  points can form a basin boundary, in discontinuous maps the line
  of discontinuity and its pre-images can also form a basin boundary.
\end{enumerate}

We have used the theoretical framework developed in this paper to explain
the non-standard bifurcation behavior of the Static VAR Compensator.
We have shown that the system yields a discontinuous map. As the
firing angle is continuously varied, there comes a parameter range
where no fixed point exists. In that situation, the state jumps
discontinuously to high periodic orbits that occur in this range of
the parameter, as per the prediction of our theory.

We believe that this body of knowledge will help in understanding the
nonstandard bifurcations observed in a number of physical systems that
yield discontinuous maps on discrete-time modeling. We also hope that
this work will provoke further investigation on the dynamics of
discontinuous maps.

\section{Acknowledgements}
One of the authors B. Rakshit is thankful to Council of Scientific and
Industrial Research, Government of India, for a junior research
fellowship.


\end{document}